# HIV, Cardiovascular Diseases, and Chronic Arsenic Exposure co-exist in a Positive Synergy

**Arghya Panigrahi[1], Amit K Chattopadhyay[2], Goutam Paul[3], Soumya Panigrahi[4]\*,**

1. Department of Physiology, Jhargram Raj College, Govt. of West Bengal. Jhargram, West Midnapore, WB, India. (apandada1@gmail.com); 2. Non-linearity and Complexity Research Group - Aston University, Aston Triangle, Birmingham, B4 7ET, UK. (a.k.chattopadhyay@aston.ac.uk); 3. Department of Physiology, University of Kalyani, West Bengal, India. (gpaul@klyuniv.ac.in) 4. Department of Medicine, Division of Infectious Diseases, Case Western Reserve University/University Hospitals of Cleveland, Cleveland, OH, 44106 (sxp579@case.edu). *corresponding author)

## Abstract

Recent epidemiological evidences indicate that arsenic exposure increases risk of atherosclerosis, cardio vascular diseases (CVD) such as hypertension, atherosclerosis, coronary artery disease (CAD) and microangiopathies in addition to the serious global health concern related to its carcinogenic effects. In experiments on animals, acute and chronic exposure to arsenic directly correlates cardiac tachyarrhythmia, and atherogenesis in a concentration and duration dependent manner. Moreover, the other effects of long-term arsenic exposure include induction of non-insulin dependent diabetes by mechanisms yet to be understood. On the other hand, there are controversial issues, gaps in knowledge, and future research priorities of accelerated incidences of CVD and mortalities in patients with HIV who are under long-term anti-retroviral therapy (ART). Although, both HIV infection itself and various components of ART initiate significant pathological alterations in the myocardium and the vasculature, simultaneous environmental exposure to arsenic which is more convincingly being recognized as a facilitator of HIV viral cycling in the infected immune cells, may contribute an additional layer of adversity in these patients. A high degree of suspicion and early screening may allow appropriate interventional guidelines to improve the quality of lives of those affected. In this mini-review which have been fortified with our own preliminary data, we will discuss some of the key current understating of chronic arsenic exposure, and its possible impact on the accelerated HIV/ART induced CVD. The review will conclude with notes on recent developments in mathematical modeling in this field that probabilistically forecast incidence prevalence as functions of aging and life style parameters, most of which vary with time themselves; this interdisciplinary approach provides a complementary kernel to conventional biology.

ABBREVIATIONS: $As_2O_3$, arsenic trioxide; AsV, arsenate; AsIII, arsenite; CAD, coronary artery disease; CVD, cardiovascular disease; EC, endothelial cell; NO, nitric oxide; NOS, nitric oxide synthase; eNOS, endothelial nitric oxide synthase; ROS, reactive oxygen species.

# Introduction

## The World of Arsenic

The knowledge of medicinal and homicidal use of arsenic can be traced back to the early days of human civilization. The poisonous and carcinogenic properties of arsenic compounds, or "Arsenicals", have been known for more than 2000 years [2, 3]. In the west, much of the current awareness can be traced back to the 1944 comedy movie *Arsenic and Old Lace* (http://www.imdb.com/title/tt0036613/). Emperors, kings, arctic explorers, heirs and commoners have been treated with arsenicals for both legitimate and homicidal purposes. A popular myth suggests that in 55 AD, the infamous Roman emperor Nero poisoned his stepbrother Britannicus with arsenic before his 14th birthday. Conversely, Hippocrates used arsenicals to treat ulcers.

> 1. In the 1800s, **Fowler's solution**, a 1% potassium arsenite solution, was used as a general tonic for treating leukemia, psoriasis, and asthma. Fowler's solution was not withdrawn from the US market until the 1950s.
> 2. The arsenic-containing drug **melarsoprol** is still the drug of choice for treating African trypanosomiasis.

> **An ACUTE FATAL DOSE** of Arsenic is in the range of 2-20 mg/kg body weight/day. Thus, a relatively healthy person with a body weight of 75 kg may die if he ingests about 145 mg to 1.45gm of Arsenic Trioxide (most commonly available arsenic compound). Considering the high density of the oxide, **less than 1/8 of a tea spoon can be fatal.** Smaller amounts may be fatal for exposed unhealthy people, elderly or children. The symptoms of poisoning by SMALL amounts of Arsenic are indistinguishable from symptoms of other illness.

> **AIR**: Arsenic in air range from 0.02 to 4ng/m$^3$) in remote and rural areas, to (3 to ~ 200ng/m$^3$) in urban areas, higher concentrations (more than 1000 ng/ m$^3$) can be found near industrial, in some countries.
>
> **WATER**: Open ocean seawater are typically low (1–2 $\mu$g/l) whereas surface waters may be 1000 times higher (up to 5000 $\mu$g/l). Arsenic levels in groundwater are typically as low as in open ocean water (about 1–2 $\mu$g/l), except in areas with volcanic rock and inland mineral deposits where arsenic levels can range up to 3000 $\mu$g/l.
>
> **In sediment:** Arsenic concentrations range from 5 to 3000 mg/kg. The higher levels are found in areas contaminated by mining and smelting. In soil, concentrations range from 1 to 40 mg/kg, usually averaging around 5 mg/kg.
>
> **Marine Biome:** Marine organisms normally contain arsenic residues ranging from < 1 to more than 100mg/kg, predominantly as organic arsenic species such as arsenosugars (macroalgae) and arsenobetaine (invertebrates and fish).
>
> **Terrestrial plants**: Plants may accumulate arsenic by root uptake from the soil or by adsorption of airborne arsenic deposited on the leaves. Arsenic levels are higher in biota collected near anthropogenic sources or in areas with geothermal activity.

In the course of time, Arsenicals made their place in medicine for treating sleeping sickness, syphilis, tuberculosis and certain skin diseases. In the early 1800s, Arsenic exposure was linked to cancer leading to a progressive diminution in its medicinal use. Arsenical have widespread use in embalming during and after the American Civil War, and today, Arsenic leaching from old cemeteries is a known groundwater pollutant in the Unites States. Arsenic has been used widely in the United States as a wood preservative until the year 2004 [4]. Eventually, as wood comes into contact with soil and water, it releases Arsenic to the environment. This is a major source of Arsenic in the environment, in particular in children playgrounds. In addition, Arsenic plays an important role in the semiconductor industry in the form of gallium arsenide, a compound used in a wide range of electronic devices [5].

> **Sources of Occupational Exposure to Arsenic**
> 1. **Semi-conductors industry**, notably of fast chips based on gallium arsenide often involving the toxic gas ARSINE, AsH$_3$.
> 2. **Glass manufacturing**.
> 3. **Mining operations and purification of metal ores and metals**. Specifically, sulfide base ore processing ( iron pyrite, lead sulfide)
> 4. **Manufacturing of certain drugs, Cattle and poultry food additives** (some anti-cancer drugs, e.g. Roxarsone).
> 5. **Manufacturing and use of pesticides** and insecticides which contain Arsenic.
> 6. **Wood Treatment**
> 7. **Rat and animal poisons**.
> 8. **Old paints**
>
> N.B. Although in most countries the use of Arsenic and its compounds for Applications 5, 6, 7 and 8 has been banned or discontinued, there are still many products that contain Arsenic which were manufactured before the ban took effect. This includes

Most Arsenic in the terrestrial environment is found in rocks and soils. Arsenic in surface and ground water is mostly a mixture of arsenite and arsenate. Arsenic is widely distributed in food; particularly high levels are found in seafood [6]. The major man-made sources of Arsenic include coal combustion, nonferrous metal smelting, and the burning of agricultural wastes. Arsenic compounds have been widely used as herbicides, fungicides, wood preservatives, desiccants, cattle and sheep dips, and as coloring agents. Arsenic continues to be widely used in agriculture, in glass and ceramics, as a metal alloy, and in semiconductors and other electronic devices. In the past, Arsenic containing rodenticides and ant poisons were responsible for many exposures. Suicidal and homicidal poisonings continue to be reported. Bacteria within soils and sediments can transform arsenate to arsenite, which can be converted into methylarsenic acid. Also within the soils and sediments, bacteria can transform methylarsenic acid into dimethylarsinic acid. All these arsenic compounds are then sublimated through rain and thereby percolate into drinking water. Molds can convert methylarsenic acid into trimethylarsine, which can then also be available within water. In addition, molds and bacteria can convert dimethylarsinic acid into both trimethylarsine and dimethylarsine in water. Once in water, trimethylarsine and dimethylarsine can volatilize into the atmosphere. Therefore, different forms of Arsenic can be found in the soil, sediments, water, atmosphere and the food chain. (Reviewed in:[7]).

Non-occupational human exposure to arsenic in the environment is primarily through the ingestion of food and water [8, 9]. Arsenic has been detected in groundwater in several countries of the world, with concentration levels exceeding the WHO drinking water guideline value of 10 µg/L (WHO) as well as the national regulatory standards (e.g. 50 µg/L in India and Bangladesh)[10, 11]. Arsenic in groundwater is often associated with geologic sources. Today, arsenic contamination of drinking water is a major worldwide public health problem. A very large number of people in India and in different parts of the world (countries like Bangladesh, Chile, Taiwan, Mexico, Thailand, Germany and parts of China) are being highly affected due to the intake of arsenic contaminated ground water [8, 10-12]. A significant fraction of the population living in the Indian districts of West Bengal (e.g. Maldaha, Murshidabad, Nadia, Bardhaman, Hooghly, 24-Parganas, and Kolkata) are also under heavy environmental exposure to arsenic and suffer from the effects of chronic arsenic toxicities initiating an alarmingly high incidence of skin, bladder and lung and also 'Blackfoot' disease (arsenic induced oxidative stress, microvascular damage and gangrene of foot)[13-16].

The daily intake of total arsenic from food and beverages is generally between 20 and 300 µg/day. Foodstuffs such as meat, poultry, dairy products and

---

**Sources of Exposure to Arsenic in Daily Life.**

1. Most drinking water contains small amounts of Arsenic. Natural water leaches small amounts of Arsenic from rocks and sand. Some industrial operations sometimes contaminate water with Arsenic.
2. Arsenic was found in items such as wine, juice, syrup, glues and pigments.
3. Arsenic is found in many foods both as organic compounds such as methyl arsines, and as inorganic arsenates and arsinates. The organic compounds are less toxic than the inorganic compounds.
4. Inorganic arsenic compounds were found in apple juice, orange and grapefruit juice, in vinegars and salad dressings, in milk and dairy products, beef, pork, poultry and in cereal.
5. Arsenic is found in most unshelled rice but also in shelled rice and in its products.

---

Roxarsone is an **organoarsenic compound** that is widely used in the poultry, pork, and cattle industry as a food additive to increase weight gain and improve feed efficiency. The drug was also approved in the United States for use in pigs. Roxarsone is marketed** as **3-Nitro** by Zoetis, a former subsidiary of Pfizer. In 2006, approximately **one million kilograms** of Roxarsone were produced in the U.S.
**Use of Roxarsone was discontinued in the US since 2011

cereals have higher levels of inorganic Arsenic [17]. Pulmonary exposure may contribute up to approximately 10 $\mu$g/day in a smoker and about 1 $\mu$g/day in a non-smoker, and more in polluted areas. The concentration of metabolites of inorganic arsenic in urine ranges from 5 to 20 $\mu$g of Arsenic/liter, but may even exceed 1000 $\mu$g/liter. In workplaces with up-to-date occupational hygiene practice, exposure generally does not exceed 10 $\mu$g/m$^3$ (8-h time-weighted average. Concentrations as low as 0.5 ppm Arsenic may trigger the initial symptoms of exposure [18].

The symptoms of small-to-moderate exposure to arsenic may appear immediately or only after several hours include headaches, vertigo and nausea. Subsequently, problems related to oxygen transfer in the body due to interferences with the hemoglobin, along with possible impairment of renal functions.

**A few important facts pertinent to Arsenic Poisoning:**

1. Every person is exposed daily to Arsenic because most drinking waters and foods contain trace amounts of arsenic.
2. The body can tolerate small doses of Arsenic without noticeable immediate physiological effects.
3. Arsenic atoms may be present in compounds in the trivalent or the pentavalent form.
4. Arsenic may be **connected to carbon** atoms, in which case the compound is called **ORGANIC** or may be connected to other types of atoms, when it will be called INORGANIC. Inorganic arsenic compounds are in general much more toxic than organic compounds.
5. Many Arsenic compounds will impart a garlic-like smell to a food or to the breath of the person who consumes them.
6. Since the chemical properties of Arsenic are very similar to the properties of phosphorous, and because phosphorous compounds are critical in numerous biochemical functions in the body, Arsenic is toxic in all functions where biochemical

Exposure to small doses of Arsenic compounds over a long period of time can result in lung, skin or liver cancer, or cancer of the lymphatic system [19], including damage to organs, such as the esophagus, as also confusion and disorientation. Arsenic mainly enters via drinking water or food that has been eaten but it can also penetrate into the body through the skin or the lungs [20].

**Overview of the human exposure to arsenic**

- **Non-occupational** human exposure to arsenic in the environment is primarily through the ingestion of food and water.
- **The daily intake of total arsenic from food and beverages is generally between 20 and 300 $\mu$g/day.**
- **Foodstuffs** such as **meat, poultry, dairy products and cereals** have higher levels of inorganic arsenic.
- **Pulmonary exposure** may contribute up to approximately **10 $\mu$g/day in a smoker** and about **1 $\mu$g/day in a non-smoker**, and more in polluted areas.
- The concentration of metabolites of inorganic arsenic in urine ranges from 5 to 20 $\mu$g of arsenic/liter, but may even exceed 1000 $\mu$g/liter.
- In workplaces with up-to-date occupational hygiene practices, exposure generally does not exceed 10 $\mu$g/m$^3$ (8-h time-weighted average.

Most Arsenic compounds are soluble in water to some extent and thus are easily transported in the blood stream and assimilated by the body. The water solubility also helps remove some of the Arsenic via the urine and the excrement. The urine and other excrements may change to darker red-brown or even greenish color [21]. Notably, a significant portion of the ingested Arsenic is absorbed by various bodily tissues and is retained over an extended period of time. Some of the Arsenic that enters the body is excreted out of the body rather rapidly but a fraction of it accumulates in various organs including the blood vessels in addition to the hair and the nails. The fraction of Arsenic that is retained in the body depends on the specific Arsenic compound that entered, the portal that it entered through, i.e. ingestion or inhalation, and other components present in the food that were consumed together. While such arsenic based toxic wastes are liberated very slowly, nevertheless they cause physiological damage in various organ systems [22].

Recent clinical research has shown that arsenic trioxide, administered intravenously, could induce cancer remission in some people with refractory acute promyelocytic leukemia [23, 24]. Although the highly suggestive but inconclusive epidemiological evidences support the Arteriosclerotic lesions, ischemic heart disease and thereby increased mortality risk was increased among arsenic-exposed persons [25, 26]. Arsenic can trigger multiple abnormal electrocardiographic patterns not limited to ventricular tachyarrhythmia [27, 28]. While the carcinogenic and cancer therapeutic potentials of arsenic have widely been studied, there is a relatively smaller attention to arsenic induced CVDs [29].

## Arsenic and Cardiovascular diseases

Being long considered a potent human health hazard due to its neoplastic outcomes, Arsenic also shows increasing epidemiological evidence of links between Arsenic exposure and risks of CAD and CVDs. As mentioned earlier, Arsenic is a major risk factor for the endemic peripheral artery disease characterized by severe arteriosclerosis and subsequent gangrene of affected extremities, so-called "Blackfoot" disease (BFD) [30, 31]. In addition, systemic vascular disease have been epidemiologically linked to chronic Arsenic exposure. High mortality from ischemic heart disease was first reported in copper smelter workers who were exposed to arsenic [32]. In the United States epidemiological studies correlations between standard mortality ratios for cardiovascular diseases and arsenic levels in food and drinking water [33, 34]. In our experimental system acute exposure of inorganic arsenic induced tachycardia in isolated amphibian hearts similar to adrenalin that could have been blocked by a beta blocker (Figure 1). Ischemic heart disease and cerebral infarction are considered late clinical manifestations of generalized atherosclerotic process and ultrasound studies in the superficial carotid artery indicated a dose response relationship between carotid atherosclerosis and chronic exposure to arsenic [33, 35, 36]. In the aforementioned studies, a significant biological links have been observed after careful adjustment of the statistical data for major CVD risk factors, signifying arsenic exposure as a potential independent risk factor for atherosclerosis, CAD, and increased incidences of cardiovascular morbidity, mortality.

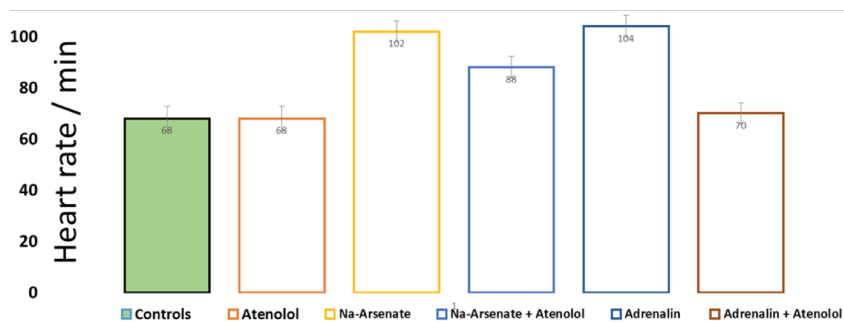

**Figure 1: Arsenic induce tachycardia in isolated amphibian heart.** Brief exposure of inorganic Arsenic (Na-Arsenate) induce tachycardia that could be counteracted by beta blockers in isolated amphibian heart.

**Chronic low dose exposure to Arsenic induce endothelial dysfunction**

The endothelium, being most sensitive to systemic or micro-environmental stress, is one of the prime target of chronic arsenic toxicity. Naturally occurring and highly toxic inorganic form arsenic rapidly induces reactive oxygen species (ROS) formation and thereby induce proliferation response in vascular smooth muscle cells. The net effect of such activity is attenuation of endothelium-dependent conduit artery dilation via downregulation of endothelial NO synthase, events that are temporally matched to the accumulation of oxidants across the vessel wall. Reduced expression of endothelial nitric oxide synthase (eNOS) leading to diminished bioavailability of vasodilating nitric oxide (NO) is a hallmark of endothelial dysfunction.

A number of earlier publications indicate that Arsenic imposes serious threats to cardiovascular health and co-morbidity. Reports of chronic arsenic exposure induced CVD related mortality, endothelial dysfunction, dysregulated lipid metabolism, and some other detailed mechanistic studies are also starting to emerge [1, 37-41]. Moreover, in a recent study it was shown that even habitual fish intake, although generally reflect healthier dietary habits with favorable effects on the endothelial function, can actually increase the chronic exposure levels of Arsenic and subsequent deleterious alteration in the arterial flow mediated dilation (FMD) – a clinical indicator of endothelial dysfunction [42]. Chronic low exposure to arsenic induce Cyclin D1 dependent NFkB/BCL3 mediated pathways leading to cellular proliferation [43]. In addition, other associated deleterious effects of chronic Arsenic exposure includes generation and persistence of proactive platelet, induction of dyslipidemia and increase endothelial adhesion of activated monocytes initiating a pro-atherogenic condition [38, 44].

Our recent studies indicate that a range of ultra-low concentrations of Arsenic (10nM – 10$\mu$M) exposure significantly down-regulate the mRNA levels of eNOS in addition significantly diminished expression of Krüppel-like Factor 2 (KLF2) and Krüppel-like Factor 4 (KLF4) in HUVEC (not shown), and in primary human aortic endothelial cells respectively. Krüppel-like factors 2 and 4 are pivotal laminar flow inducible transcription factors that modulate several genes critical for maintaining an antithrombotic endothelial surface [45, 46]. KLF2 is a key determinant of the anti-inflammatory and anti-atherogenic vascular environment [47-49], and a potent inducer of endothelial nitric oxide synthase (eNOS) [50]. KLF2 blocks expression of the procoagulant Tissue Factor [51] and also inhibits thrombin-induced activation of endothelial cells by decreasing expression of the thrombin-activated protease activated receptor type 1 (PAR-1), conversely, TNF-α, IL-1β, and oxidative stress, repress KLF2 expression in the endothelial cells (EC) [52]. KLF4 is also an independent regulator of EC function, and is protective against atherosclerosis [53]. According to reports, Krüppel-like factors (KLFs) are important mediators of monocyte differentiation and activation as well. KLF2 is a negative regulator of proinflammatory genes both in the ECs and monocytes and expression of KLF2 in monocytes is reduced following exposure to TNF-α or LPS and in monocytes from patients with coronary artery disease [54]. Krüppel-like factor 4 is also involved in myeloid cell differentiation [55] and is a critical regulator of macrophage polarization; in vitro exposure of mouse macrophages to LPS, proinflammatory cytokines, or oxLDL, results in decreased KLF4 levels [53, 56]. Mouse macrophages deficient in KLF4 have enhanced foam cell formation in response to oxLDL [53]. Our preliminary observations shows brief exposure of inorganic Arsenic in nanomolar concentration induce significant down-regulation of several other endothelial regulator genes including increased expression of VCAM-1, and E-selectin ( unpublished observations, not shown) in HUVEC (Human Umbilical Vein Endothelial Cell).

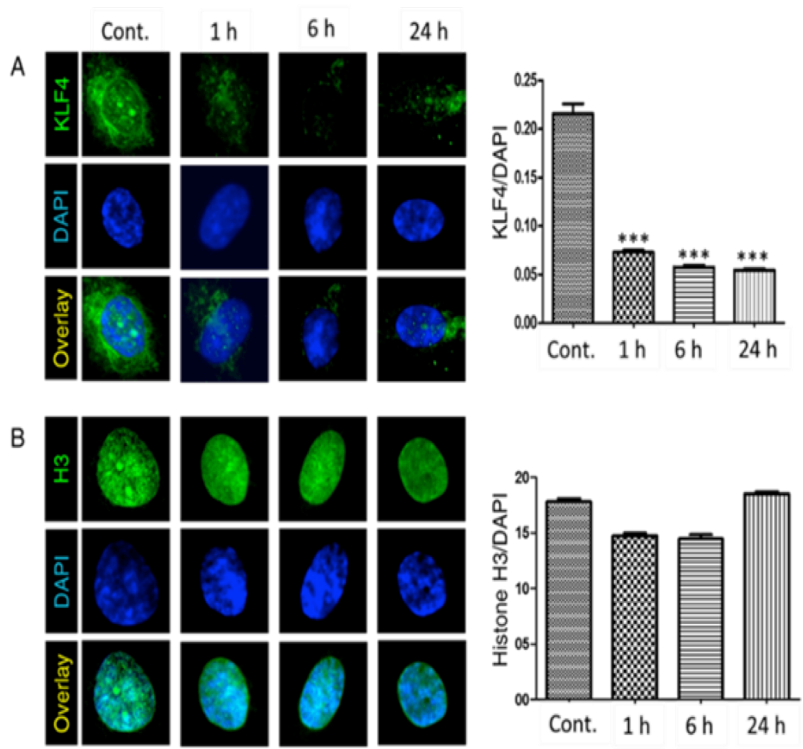

To determine whether the reduced eNOS expression in Arsenic exposed endothelial cells and other peripheral blood mononuclear cells (not shown), could be linked to a reduced expression of endothelial KLF4 (and KLF2: not shown), we examined nuclear KLF4 levels by immunofluorescence microscopy, normalizing the digitized nuclear KLF4 fluorescence to nuclear DNA (as measured by DAPI fluorescence). We observed significantly reduced KLF4 expression in Arsenic exposed primary human aortic endothelial cells when compared to the KLF4 expression in untreated control cells (Figure 2a. ***p < 0.001). Interestingly, when the nuclear levels of histone H3, as an internal control, was measured using identical parameters no such dose or time dependent effect of Arsenic exposure on in the endothelial cells was observed (Figure 2b.). We confer, the vascular micro-environment during Arsenic exposure is pathologically modified characterizing a diminished expression of the anti-inflammatory and anti-atherogenic transcriptional regulator KLF2 and KLF4 in addition to significant reduction in eNOS production and increased expression of endothelial adhesion molecules, like VCAM-1 and E-selectin. Here, we provide a comprehensive set of data showing the effects of low dose Arsenic exposure (10$\mu$M) at different time point on cultured primary human aortic endothelium from descending thoracic aorta (Figure 2).

**Figure 2: Arsenic exposure reduce Krüppel-like Factor 4 expression in human aortic endothelial cells in a time and dose dependent manner.** Human primary aortic endothelial cells were grown in sterile chamber slides, and exposed to 10mM of Arsenic at the indicated time points. The cells were subsequently immunostained and digital images were recorded by epi-fluorescence microscopy (EVOSÒFL, Life Technologies) using a 100X objective (20X images for large scale data acquisition and analyses) for the detection and analysis of nuclear KLF4 (Panel a) and Histone H3 (Panel b.) as internal control. Quantitative fluorescence intensity ratio (KLF4:DAPI) data of 100 cells at each time point from 3 independent experiments are presented to the right of each panel. The data were compared using a two tailed Mann Whitney test. ***p<0.001.

Arsenic-induced molecular and cellular events related to atherogenesis: In *vitro studies* cultured human endothelial cell indicate that arsenic can initiate oxidative damage, activation of transcription factors, and gene expression relevant to endothelial dysfunction and CVDs (Figure 3). Chronic exposure to Arsenic is also a potential risk factor for type 2 diabetes.

Exposure to inorganic Arsenic induces pre-diabetic effects by altering deregulating lipid metabolism, gluconeogenesis and insulin secretion in healthy individuals, and worsens glucose metabolism in established diabetics although insulin resistance might be not the sole reason of diabetic effects caused by inorganic Arsenic [57].

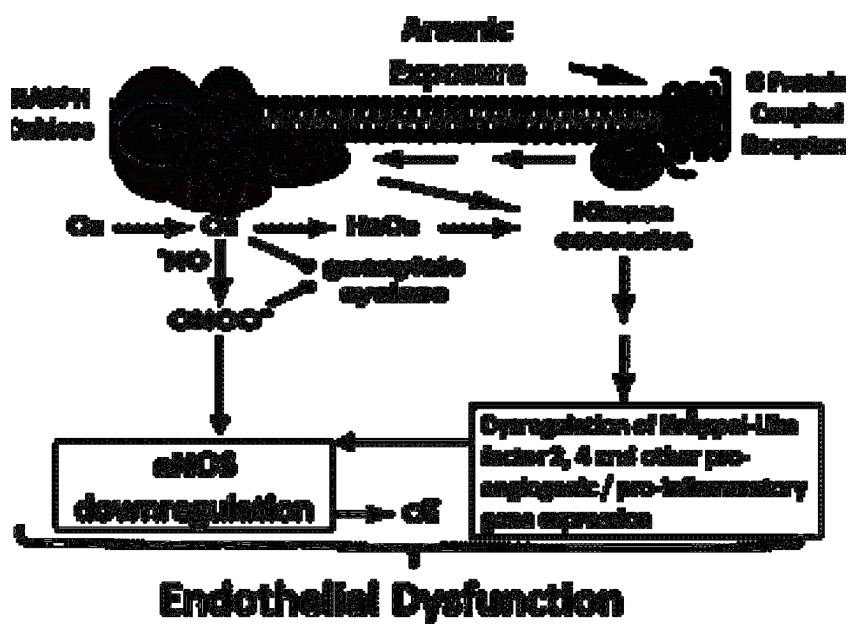

**Figure 3: Schematic view of plausible arsenic induced oxidative damage and endothelial dysfunction.** Arsenic interacts with G Protein Coupled Receptors (GPCR) to initiate signal amplification schemes regulating NOX-dependent redox signaling. Downstream signaling for dysfunction was (partially adapted from: States JC, et al., Toxicol Sci. 2009.[1]).

**Arsenic exposure, cardiovascular diseases and global HIV incidences:** Cardiovascular abnormalities are common in HIV-infected individuals but often go unrecognized or untreated, which results in increased cardiovascular-related morbidity and mortality and reduced quality of life. Clinicians may mistakenly attribute signs of cardiovascular abnormalities to pulmonary or infectious causes, an error that can delay appropriate treatment. Despite a dramatic improvement in survival in the antiretroviral therapy (ART) era, there remains an increased risk of thromboembolic and cardiovascular co-morbidities in these treated [58-60]. The core mechanism(s) that contribute to this increased risk of CVD in HIV disease have not been fully elucidated, but may be partially related to chronic immune activation [61, 62]. Systemic indices of inflammation and coagulation have been linked to cardiovascular risk in HIV infection [63] and PET CT scans have demonstrated metabolic evidence of aortic inflammation in treated HIV infection [64].

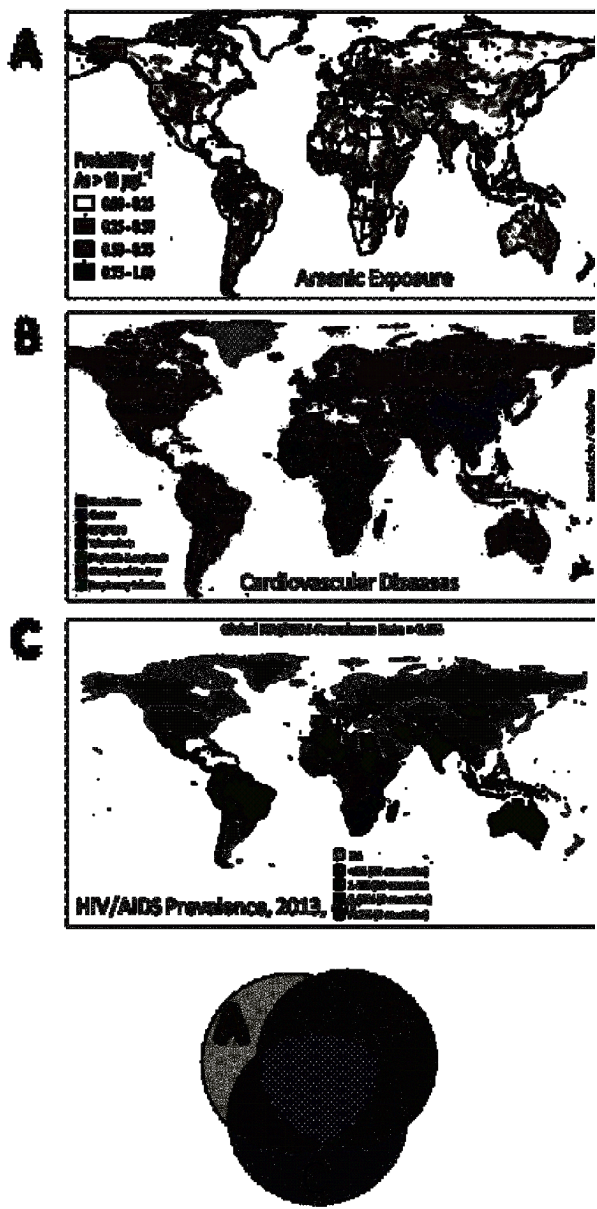

**Figure 4:** Global Arsenic exposure (A), Cardiovascular/other diseases (B), HIV/AIDS Prevalence in 2013 (C), and the projected overlaps (lower panel).

Endothelium is a highly reactive surface, responding to a myriad of micro-environmental stress-inducing factors resulting from infection tissue injury, and inflammation [65]. ECs serve as selective micro environmental barriers for the exchange of fluid and macromolecules from the vascular compartment to the tissue. In addition, the bidirectional relationship that follows between the coagulation pathways and vascular inflammation modulating endothelial homeostasis is widely recognized [66, 67].

We showed in one of our recent reports that eNOS and KLF2 are significantly downregulated inducing endothelial dysfunction along with a dramatic increase in the subendothelial migration of CD8 T cells and inflammatory monocytes in the aortic endothelium of *Rhesus macaque* who are acutely infected with simian immunodeficiency virus (SIV)- the HIV homologue infecting non-human primates [68] How such downregulation affects the formation and sustenance of the immunological synapse bond is of extreme importance in defining the immune lifeline of an affected individual. An immediate future extension of our present line of research in this field is to analyze such effects based on mathematical and probabilistic models that will also highlight the quantitative importance of all affecting factors.

Chronic low dose exposure to inorganic Arsenic and Arsenicals, endothelial dysfunction, CVD, and an association of endemic HIV infections opens up future avenues of research into, what is a relatively poorly understood topic with major implications for human health. While the data acquisition on the world prevalence of HIV and cardiovascular diseases is thorough, the data on chronic exposure to Arsenic is still incomplete for most part of the world. But, we know a large sum of the world population, mostly unknowingly, is exposed to Arsenic irrespective of the socioeconomic status or geographical location of their country. This is a fact which generally indicate that the HIV infected population who are on continued anti-retroviral therapy (ART) with controlled viremia, are also exposed to low, moderate or high levels of environmental Arsenic that might be an additional trigger in initiating their accelerated endothelial dysfunction and atherosclerotic lesions. Determination of the true risk of CVD in HIV and measuring the synergistic effects of Arsenic exposure that further complicates the pathogenesis of CVD will become increasingly important in the long-term management of HIV-infected patients receiving antiretroviral therapy.

**Arsenic Modelling and Transport**

As described above, Arsenic is now widely accepted to have a dual role as a causative and a curative agent primarily in non-infectious disease proliferation [69] but also as a secondary instigator of population infection [70]. While conventional biology is typically a first resort, probabilistic prediction of affectation and disease proliferation often stretch beyond the realms of experimental data, especially in predicting what levels of fluctuation of the affecting parameters could instigate a certain threshold of arsenic inflicted persecution. As an example, if the water arsenic toxicity factor increases by 5% over the next 3 years while the glutathione depletion factor decreases by 7% leading to an overall protein inactivation increase by 8% (considering only a few token factors) during the same time span, what will be the increase in arsenic afflicted atherosclerosis and CVD during this time? To address such questions in a non-invasive procedure, computer based data analysis and statistical modeling is the new genre of science in this interdisciplinary arena.

Techniques from statistical mechanics, thermodynamics and nonlinear mechanics, together with native computer simulation, has started unearthing lots of features that have so far remained outside the conventional research purview. In a recent ensemble modeling approach, a mathematical model has been studied that predicts intracellular arsenic flow rates under varying mutant conditions, focusing on the toxicity mechanism of arsenic. This data based analyzes clearly highlight an array of important conclusions like a) protein based arsenic profiles have short-termed exposure while glutathione based arsenic compounds have longer termed ramifications; b) arsenic dynamics is mostly a transient mechanism that probabilistically escapes the vacuole via natural export mechanism, etc. While this work was studied on a yeast substrate, more indicative suggestions towards the protein-versus-glutathione debate have been advised. While this line of modeling directly assesses the microbiological origin of its affectation, secondary infections in the form of (Buruli) ulcerous swellings, believed to be arsenic inflicted, have recently been studied using data from chosen African countries [70]. The results from this study indicate without confirming as much the importance of environmental implication of arsenic disease propagation. Similar results were obtained a while ago by a South African modeling group who studied the optimal control of arsenic transmission dynamics targeting *mycobacterium ulceran* infection [71]. While the data analysis results [70] indicated the roots of the proliferation mode, this more recent kinetic equation based nonlinear modeling [71] traced the mitigation mechanism as an optimal combination of environmental and health education together with water purification protocols.

Modeling attempts at arsenic retention processes in wetland areas is another line of complementary research that focuses on the role of trace elements in arsenic effect proliferation. Analyzing a model representing arsenic propagation dynamics in constructed wetlands targeting a quantitative prediction of the mutual relationship between iron and arsenic retention efficiency increase, the model results indicate a maximum arsenic retention efficiency of 85-95% [72]. What all such effects ultimately culminate at is in the population response to exposure of chemical carcinogens at low to high exposure levels often leading to a mutation of a single-cell DNA, hastening a terminal illness. A consummate detailing of such "carcinogenesis modeling" can be availed from a self-sufficient overview by Vineis, et al. [73]. Of the five models discussed here, only Model 4 indirectly alludes to arsenic affectation of carcinoma, another indication of insufficient research in this field.

The above highlighted areas are only the tips of the proverbial iceberg. More concerted efforts combining complementary inter- and cross- disciplinary efforts are needed to tackle this problem, an approach that has taken pace off late.

## Concluding Remarks

Studies summarized in this mini-review suggest that chronic Arsenic exposure induces pathophysiological events relevant to the atherogenic potential including impaired vascular nitric oxide homeostasis, low Krüppel-like factor 2, 4 expression, and enhanced expression endothelial adhesion molecule like VCAM-1. These detrimental changes in the endothelial microenvironment, collectively expressed as endothelial dysfunction are the key molecular events in the cardiovascular system which are surprisingly common to Arsenic exposure and in the HIV infected patients on long-term ART. Based on the accumulating epidemiological evidences and experimental data, we conclude, arsenic exposure could be considered as an important modifying factor in the understating and intervention strategies of atherosclerosis and related cardiovascular diseases in virologically controlled HIV patients on ART, at least under certain circumstances (including genetic background, diet, co-exposure and geographical location). Also, we would argue that the considerable overlap with the worldwide exposure of Arsenic, incidences of HIV infection positively reinforce the initiation, persistence, and progress of HIV infection itself and the pathogenesis of related cardiovascular complications. While mathematical modeling of arsenic poisoning, infection and its impact on carcinogenesis is a vital and newly adopted line of research, the sheer paucity of material in gelling together known biological observations against data and mathematical models clearly indicates the need for more detailed and target driven studies. Over the next couple of years, such cross-platform research is expected to hold center stage in leading the light in this field.

**Conflict of Interest Statement:**
All other authors declare that they have no competing interests.

**Authors' contribution:**

AP and SP designed, performed, analyzed experiments, and SP, AKC, and AP wrote the manuscript with collaborations from GP, and AKC. All authors contributed to general design and discussion of the project and reviewed and approved the manuscript.

**Acknowledgements:**
SP acknowledge the Fasenmyer Foundation; the CWRU Center for AIDS Research; the Cleveland Immunology Consortium (CLIC) for supporting this study without having any role in study design, data collection or analysis, the preparation of the manuscript or the decision to publish it.

# REFERENCES:

1. States, J.C., et al., *Arsenic and cardiovascular disease.* Toxicol Sci, 2009. **107**(2): p. 312-23.
2. Bhattacharya, P., et al., *Arsenic in the environment: Biology and Chemistry.* Sci Total Environ, 2007. **379**(2-3): p. 109-20.
3. Naidu, R. and P. Bhattacharya, *Arsenic in the environment--risks and management strategies.* Environ Geochem Health, 2009. **31 Suppl 1**: p. 1-8.
4. Sibbald, B., *Arsenic and pressure-treated wood: the argument moves to the playground.* CMAJ, 2002. **166**(1): p. 79.
5. Chen, H.W., *Gallium, indium, and arsenic pollution of groundwater from a semiconductor manufacturing area of Taiwan.* Bull Environ Contam Toxicol, 2006. **77**(2): p. 289-96.
6. Ma, J., *[Determination of inorganic arsenic in sea food].* Zhonghua Yu Fang Yi Xue Za Zhi, 1984. **18**(3): p. 178-9.
7. Mandal, B.K. and K.T. Suzuki, *Arsenic round the world: a review.* Talanta, 2002. **58**(1): p. 201-35.
8. Samal, A.C., et al., *Human exposure to arsenic through foodstuffs cultivated using arsenic contaminated groundwater in areas of West Bengal, India.* J Environ Sci Health A Tox Hazard Subst Environ Eng, 2011. **46**(11): p. 1259-65.
9. Cwiek, K., *[Arsenic in food and its transformation in the environment].* Rocz Panstw Zakl Hig, 1988. **39**(6): p. 438-49.
10. Mukherjee, A., P. Bhattacharya, and A.E. Fryar, *Arsenic and other toxic elements in surface and groundwater systems.* Applied Geochemistry, 2011. **26**(4): p. 415-420.
11. Hoque, M.A., et al., *Delineating low-arsenic groundwater environments in the Bengal Aquifer System, Bangladesh.* Applied Geochemistry, 2011. **26**(4): p. 614-623.
12. Zaldivar, R., *Arsenic Contamination of Drinking-Water and Foodstuffs Causing Endemic Chronic Poisoning.* Beitrage Zur Pathologie, 1974. **151**(4): p. 384-400.
13. Chakraborty, A.K. and K.C. Saha, *Arsenical dermatosis from tubewell water in West Bengal.* Indian J Med Res, 1987. **85**: p. 326-34.
14. Mazumder, D.N., et al., *Arsenic contamination of ground water and its health impact on population of district of nadia, west bengal, India.* Indian J Community Med, 2010. **35**(2): p. 331-8.
15. Jomova, K., et al., *Arsenic: toxicity, oxidative stress and human disease.* J Appl Toxicol, 2011. **31**(2): p. 95-107.
16. Flora, S.J., et al., *Arsenic induced oxidative stress and the role of antioxidant supplementation during chelation: a review.* J Environ Biol, 2007. **28**(2 Suppl): p. 333-47.
17. Sullivan, T.W. and A.A. al-Timimi, *Safety and toxicity of dietary organic arsenicals relative to performance of young turkeys. 4. Roxarsone.* Poult Sci, 1972. **51**(5): p. 1641-4.
18. Abernathy, C., *EXPOSURE AND HEALTH EFFECTS.* 2001.
19. Martinez, V.D., et al., *Arsenic exposure and the induction of human cancers.* J Toxicol, 2011. **2011**: p. 431287.
20. Kundu, M., et al., *Precancerous and non-cancer disease endpoints of chronic arsenic exposure: the level of chromosomal damage and XRCC3 T241M polymorphism.* Mutat Res, 2011. **706**(1-2): p. 7-12.

21.     Ishinishi, N., et al., *[Symptoms and diagnosis of poisoning: arsenic and arsenic compounds (metalloid)].* Nihon Rinsho, 1973. **31**(6): p. 1991-9.

22.     Suzuki, K.T., B.K. Mandal, and Y. Ogra, *Speciation of arsenic in body fluids.* Talanta, 2002. **58**(1): p. 111-9.

23.     Douer, D. and M.S. Tallman, *Arsenic trioxide: new clinical experience with an old medication in hematologic malignancies.* J Clin Oncol, 2005. **23**(10): p. 2396-410.

24.     Leu, L. and L. Mohassel, *Arsenic trioxide as first-line treatment for acute promyelocytic leukemia.* Am J Health Syst Pharm, 2009. **66**(21): p. 1913-8.

25.     Chen, C.J., et al., *Atherogenicity and carcinogenicity of high-arsenic artesian well water. Multiple risk factors and related malignant neoplasms of blackfoot disease.* Arteriosclerosis, 1988. **8**(5): p. 452-60.

26.     Zaldivar, R. and G.L. Ghai, *Mathematical model of mean age, mean arsenic dietary dose and age-specific prevalence rate from endemic chronic arsenic poisoning: a human toxicology study.* Zentralbl Bakteriol B, 1980. **170**(5-6): p. 402-8.

27.     Hall, J.C. and R. Harruff, *Fatal cardiac arrhythmia in a patient with interstitial myocarditis related to chronic arsenic poisoning.* South Med J, 1989. **82**(12): p. 1557-60.

28.     Manna, P., M. Sinha, and P.C. Sil, *Arsenic-induced oxidative myocardial injury: protective role of arjunolic acid.* Arch Toxicol, 2008. **82**(3): p. 137-49.

29.     Benbrahim-Tallaa, L., et al., *Molecular events associated with arsenic-induced malignant transformation of human prostatic epithelial cells: aberrant genomic DNA methylation and K-ras oncogene activation.* Toxicol Appl Pharmacol, 2005. **206**(3): p. 288-98.

30.     Tseng, W.P., *Blackfoot disease in Taiwan: a 30-year follow-up study.* Angiology, 1989. **40**(6): p. 547-58.

31.     Tseng, C.H., *An overview on peripheral vascular disease in blackfoot disease-hyperendemic villages in Taiwan.* Angiology, 2002. **53**(5): p. 529-37.

32.     Lee, A.M. and J.F. Fraumeni, Jr., *Arsenic and respiratory cancer in man: an occupational study.* J Natl Cancer Inst, 1969. **42**(6): p. 1045-52.

33.     James, K.A., et al., *Association between Lifetime Exposure to Inorganic Arsenic in Drinking Water and Coronary Heart Disease in Colorado Residents.* Environmental Health Perspectives, 2015. **123**(2): p. 128-134.

34.     James, K.A., et al., *Response to "Comment on 'Association between Lifetime Exposure to Inorganic Arsenic in Drinking Water and Coronary Heart Disease in Colorado Residents' ".* Environmental Health Perspectives, 2015. **123**(7): p. A169-A169.

35.     Chiou, H.Y., et al., *Dose-response relationship between prevalence of cerebrovascular disease and ingested inorganic arsenic.* Stroke, 1997. **28**(9): p. 1717-1723.

36.     Wang, C.H., et al., *Biological gradient between long-term arsenic exposure and carotid atherosclerosis.* Circulation, 2002. **105**(15): p. 1804-1809.

37.     Alissa, E.M. and G.A. Ferns, *Heavy metal poisoning and cardiovascular disease.* J Toxicol, 2011. **2011**: p. 870125.

38.     Bae, O.N., et al., *Arsenite-enhanced procoagulant activity through phosphatidylserine exposure in platelets.* Chemical Research in Toxicology, 2007. **20**(12): p. 1760-1768.

39.     Liao, Y.T., et al., *Elevated lactate dehydrogenase activity and increased cardiovascular mortality in the arsenic-endemic areas of southwestern Taiwan.* Toxicol Appl Pharmacol, 2012. **262**(3): p. 232-7.

40.     Simeonova, P.P. and M.I. Luster, *Arsenic and atherosclerosis.* Toxicol Appl Pharmacol, 2004. **198**(3): p. 444-9.

41.     Kumagai, Y. and J. Pi, *Molecular basis for arsenic-induced alteration in nitric oxide production and oxidative stress: implication of endothelial dysfunction.* Toxicol Appl Pharmacol, 2004. **198**(3): p. 450-7.

42.     Buscemi, S., et al., *Endothelial function and serum concentration of toxic metals in frequent consumers of fish.* PLoS One, 2014. **9**(11): p. e112478.

43.     Ouyang, W., et al., *Cyclin D1 induction through IkappaB kinase beta/nuclear factor-kappaB pathway is responsible for arsenite-induced increased cell cycle G1-S phase transition in human keratinocytes.* Cancer Res, 2005. **65**(20): p. 9287-93.

44.     Lemaire, M., et al., *Arsenic Exposure Increases Monocyte Adhesion to the Vascular Endothelium, a Pro-Atherogenic Mechanism.* PLoS One, 2015. **10**(9): p. e0136592.

45.     Parmar, K.M., et al., *Integration of flow-dependent endothelial phenotypes by Kruppel-like factor 2.* J Clin Invest, 2006. **116**(1): p. 49-58.

46.     Hamik, A., et al., *Kruppel-like factor 4 regulates endothelial inflammation.* J Biol Chem, 2007. **282**(18): p. 13769-79.

47.     Nayak, L., et al., *Kruppel-like factor 2 is a transcriptional regulator of chronic and acute inflammation.* Am J Pathol, 2013. **182**(5): p. 1696-704.

48.     Atkins, G.B., et al., *Hemizygous deficiency of Kruppel-like factor 2 augments experimental atherosclerosis.* Circ Res, 2008. **103**(7): p. 690-3.

49.     Mahabeleshwar, G.H., et al., *A myeloid hypoxia-inducible factor 1alpha-Kruppel-like factor 2 pathway regulates gram-positive endotoxin-mediated sepsis.* J Biol Chem, 2012. **287**(2): p. 1448-57.

50.     SenBanerjee, S., et al., *KLF2 Is a novel transcriptional regulator of endothelial proinflammatory activation.* J Exp Med, 2004. **199**(10): p. 1305-15.

51.     Lin, Z., et al., *Kruppel-like factor 2 (KLF2) regulates endothelial thrombotic function.* Circ Res, 2005. **96**(5): p. e48-57.

52.     Lin, Z., et al., *Kruppel-like factor 2 inhibits protease activated receptor-1 expression and thrombin-mediated endothelial activation.* Arterioscler Thromb Vasc Biol, 2006. **26**(5): p. 1185-9.

53.     Sharma, N., et al., *Myeloid Kruppel-like factor 4 deficiency augments atherogenesis in ApoE-/- mice--brief report.* Arterioscler Thromb Vasc Biol, 2012. **32**(12): p. 2836-8.

54.     Das, H., et al., *Kruppel-like factor 2 (KLF2) regulates proinflammatory activation of monocytes.* Proc Natl Acad Sci U S A, 2006. **103**(17): p. 6653-8.

55.     Feinberg, M.W., et al., *The Kruppel-like factor KLF4 is a critical regulator of monocyte differentiation.* Embo J, 2007. **26**(18): p. 4138-48.

56.     Liao, X., et al., *Kruppel-like factor 4 regulates macrophage polarization.* J Clin Invest, 2011. **121**(7): p. 2736-49.

57.     Maull, E.A., et al., *Evaluation of the Association between Arsenic and Diabetes: A National Toxicology Program Workshop Review.* Environmental Health Perspectives, 2012. **120**(12): p. 1658-1670.

58.     De Socio, G.V., et al., *Identifying HIV patients with an unfavorable cardiovascular risk profile in the clinical practice: results from the SIMONE study.* J Infect, 2008. **57**(1): p. 33-40.

59.     Triant, V.A. and S.K. Grinspoon, *Vascular dysfunction and cardiovascular complications.* Curr Opin HIV AIDS, 2007. **2**(4): p. 299-304.

60.     Stein, J.H. and P.Y. Hsue, *Inflammation, immune activation, and CVD risk in individuals with HIV infection.* JAMA, 2012. **308**(4): p. 405-6.

61.     Funderburg, N.T. and M.M. Lederman, *Coagulation and morbidity in treated HIV infection.* Thromb Res, 2014. **133 Suppl 1**: p. S21-4.

62.     Appay, V. and D. Sauce, *Immune activation and inflammation in HIV-1 infection: causes and consequences.* J Pathol, 2008. **214**(2): p. 231-41.

63.     Funderburg, N.T., et al., *Increased tissue factor expression on circulating monocytes in chronic HIV infection: relationship to in vivo coagulation and immune activation.* Blood, 2010. **115**(2): p. 161-7.

64.     Subramanian, S., et al., *Arterial inflammation in patients with HIV.* JAMA, 2012. **308**(4): p. 379-86.

65.     Franses, J.W. and E.R. Edelman, *The evolution of endothelial regulatory paradigms in cancer biology and vascular repair.* Cancer Res, 2011. **71**(24): p. 7339-44.

66.     Schulz, C., B. Engelmann, and S. Massberg, *Crossroads of coagulation and innate immunity: the case of deep vein thrombosis.* J Thromb Haemost, 2013. **11 Suppl 1**: p. 233-41.

67.     Levi, M., T. van der Poll, and H.R. Buller, *Bidirectional relation between inflammation and coagulation.* Circulation, 2004. **109**(22): p. 2698-704.

68.     Panigrahi, S., et al., *SIV/SHIV Infection Triggers Vascular Inflammation, Diminished Expression of Krüppel-like factor 2 (KLF2) and Endothelial Dysfunction.* J Infectious Disease, 2015((In Press)).

69.     Talemi, S. R., et al., *Mathematical modelling of arsenic transport, distribution and detoxification processes in yeast.* Molecular Microbiology **92**(6), 1343-1356 (2014).

70.     Bonyah, E., et al. *Theoretical Model for the Transmission Dynamics of the Buruli Cancer with Saturated Treatment.* Computational and Mathematical Methods in Medicine, 2014: 576039 (2014).

71.     Kimaro, M. A., et al. *Modelling the Optimal Control of Transmission Dynamics of Mycobacterium ulceran Infection.* Open Journal of Epidemiology **5**, 229-243 (2015).

72.    Llorens, E., et al. *Modelling of arsenic retention in constructed wetlands.* Bioresource Technology **147**, 221-227 (2013).

73.    Vineis, P., Schatzkin, A. and Potter, J. D. *Models of carcinogenesis: an overview.* Carcinogenesis **31**(10), 1703-1709 (2010).